\documentclass[amsmath,amssymb,aps,prc,twocolumn,superscriptaddress]{revtex4-1}

\usepackage{graphicx}
\usepackage{dcolumn}
\usepackage{bm}
\usepackage{enumitem}
\usepackage{textcomp}
\usepackage{gensymb}

\usepackage{soul}

\newcommand{\lpc}{Normandie Universit\'e, ENSICAEN, UNICAEN, CNRS/IN2P3, LPC Caen, 14000 Caen, France}
\newcommand{\cea}{Universit{\'e} Paris-Saclay, CEA, List, Laboratoire National Henri Becquerel (LNE-LNHB), F-91120 Palaiseau, France}
\newcommand{\msu}{National Superconducting Cyclotron Laboratory and Department of Physics and Astronomy,
Michigan State University, East Lansing 48824 MI, USA}
\newcommand{\ganil}{GANIL, CEA/DRF-CNRS/IN2P3, Bd Henri Becquerel, 14076 Caen, France}
\newcommand{\desy}{Deutsches Elektronen-Synchrotron DESY, Notkestra{\ss}e 85, 22607 Hamburg, Germany}
\usepackage{xcolor}
\usepackage{amsmath}
\usepackage{physics}
\usepackage{bm}

\begin{document}

\preprint{He6_halflife_pp}

\title{High precision measurement of the $^6$He half-life}

\affiliation{\lpc}
\affiliation{\msu}
\affiliation{\cea}
\affiliation{\desy}
\affiliation{\ganil}

\author{M.~Kanafani}
\affiliation{\lpc}
\author{X.~Fl\'echard}
\email{flechard@lpccaen.in2p3.fr}
\affiliation{\lpc}
\author{O.~Naviliat-Cuncic}
\affiliation{\lpc}\affiliation{\msu}
\author{G.D.~Chung}
\affiliation{\lpc}
\author{S.~Leblond}
\affiliation{\cea}
\author{E.~Li\'enard}
\affiliation{\lpc}
\author{X.~Mougeot}
\affiliation{\cea}
\author{G.~Qu\'em\'ener}
\affiliation{\lpc}
\author{A.~Simancas Di Filippo}
\affiliation{\lpc}\affiliation{\desy}
\author{J-C.~Thomas}
\affiliation{\ganil}

\date{\today}
\begin{abstract}
The half-life of $^{6}$He has been measured using a low energy 
radioactive beam implanted in a YAP scintillator and recording decay events in
a 4$\pi$ geometry. Events were time-stamped with a digital data acquisition system
enabling a reliable control of dead-time effects and detector gain variations.
The result, $T_{1/2} = (807.25 \pm 0.16_{\rm stat} \pm 0.11_{\rm sys}$)~ms, provides the most precise value obtained so far and is consistent with the only previous measurement
having a precision smaller than 0.1\%. This resolves the longstanding discrepancy previously observed between two sets of measurements.
\end{abstract}
\maketitle
\section{Introduction}
\label{sec:intro}

The study of the $^6$He beta decay has been instrumental in establishing the {\em V-A}
character of the weak interaction \cite{Joh63,Vis63}.
The simplicity of the decay has regained considerable attention in the past decades as a sensitive window to probe the electroweak standard model in the sector involving the lightest quarks. Experiments using trapped $^6$He$^+$ ions \cite{Fle11} or atoms \cite{Mue22} have been performed to address the beta-neutrino angular correlation and ancillary measurements have been carried out to study atomic effects following beta decay \cite{Cou12,Hon17}. New techniques have been proposed to determine the Fierz interference term from the beta energy spectrum \cite{Nav16,Huy18,Cres20} and selected observables of the transition have been the object of detailed theoretical studies to assess the sensitivity of future experiments to new physics \cite{Gli21}.

One of the properties that makes the $^6$He decay attractive is the fact that the transition to the $^6$Li ground-state is pure Gamow-Teller (GT). As a result, the correlation terms in the decay rate distribution or the shape of the beta energy spectrum are, to first order, independent of the dominant nuclear matrix element, noted $c$ in Ref.~\cite{Cal75}. However, due to induced form factors in the hadronic weak currents, this matrix element enters recoil order corrections in ratios with the weak magnetism form factor \cite{Cal75}. These corrections are important for instance in the extraction of the Fierz interference term from measurements of the shape of the beta-energy spectrum.

The value of $c$ extracted from Ref.~\cite{Kne12b} has a relative uncertainty of $3.6\times 10^{-4}$ so that taking only this value would result in a negligible impact for the extraction of the Fierz term from a measurement of the spectrum. However, as pointed out in Ref.~\cite{Kne12b}, the prior internal consistency of the $^6$He half-lives, having an uncertainty smaller than 1\% is very poor. A fit to the five most precise values prior to 2012 (Fig.~\ref{fig:t12-vs-year}) gives $T_{1/2} = 800.6\pm0.8$ with $\chi^2/\nu = 6.95$ with $\nu = 4$. Resolving this discrepancy requires a new measurement with comparable precision than the one reported in Refs.~\cite{Kne12b} and preferably with a technique having different sensitivity to the main sources of systematic effects.

We report here a high precision measurement of the $^6$He half-life using a digital data acquisition system which provides a time stamp when recording the energy of the beta particle. The setup includes an $^{241}$Am source which is constantly monitored in order to control gain variations.

\section{Experimental setup}
\label{sec:setup}

\subsection{Beam transport and time sequence}
\begin{figure}[ht!]
\includegraphics[width = 1.\columnwidth]{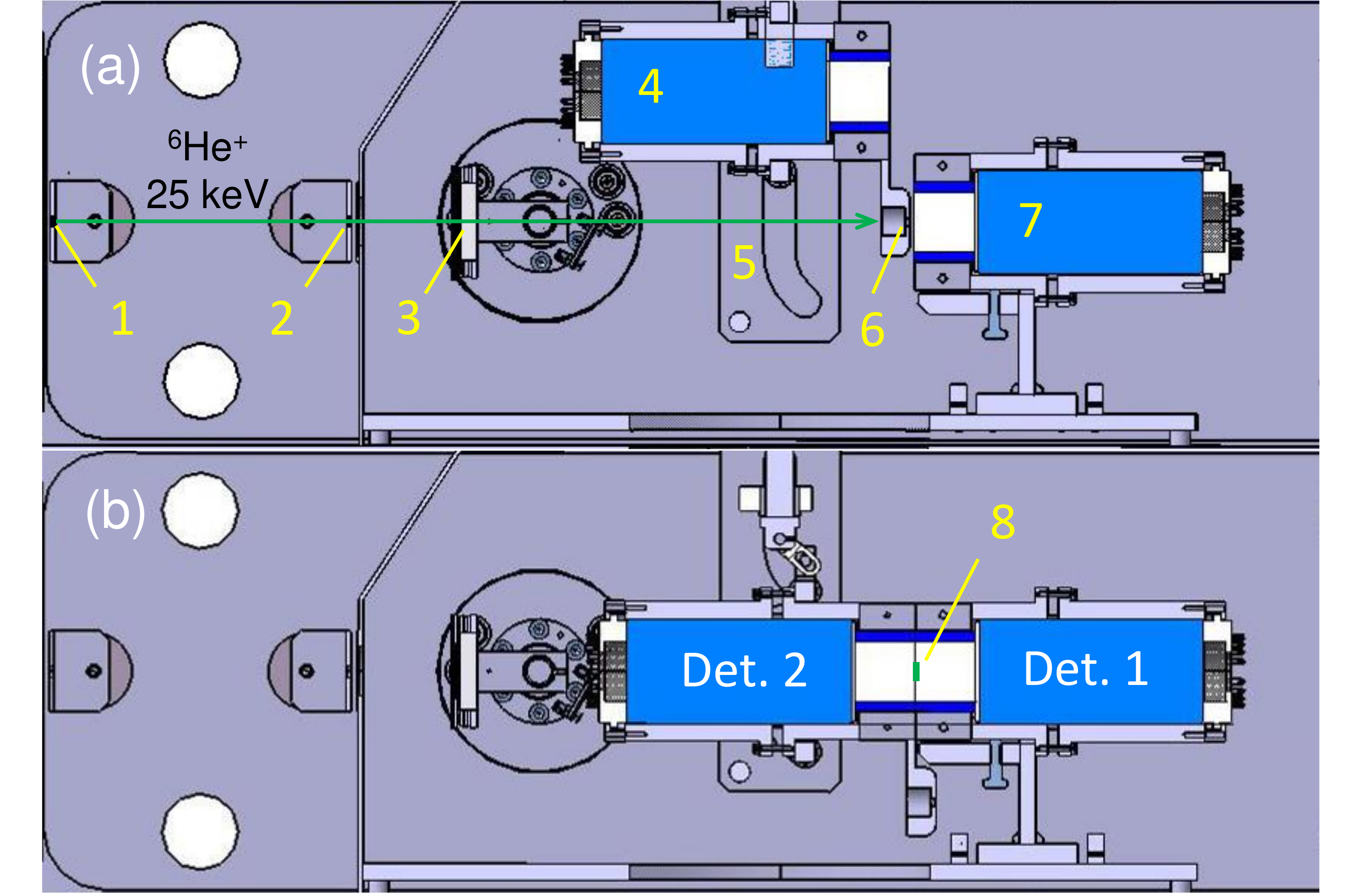}
\caption{\label{fig:setup}{Sectional view of the vacuum chamber for the implantation (a) and data taking (b) configurations. The labels on panel (a) are: 1 and 2-- the two \O 6~mm collimators in the first section of the chamber; 3-- movable Si detector; 4 and 5-- the moving detector and its mechanical guide; 6-- the third \O 4~mm collimator; 7-- the fixed detector. The green arrow indicates the $^6$He$^+$ beam. On panel (b), label 8 indicates the implantation region.}}
\end{figure}

The experiment was performed at the Grand Acc\'el\'erateur National d'Ions Lourds (GANIL), Caen. 
The $^6$He$^+$ ions were produced by the SPIRAL target-ECR ion source system
and were guided at 25~keV to the low energy beam line LIRAT after mass separation by a dipole magnet. The beam was chopped with a fast electrostatic deflector located upstream from the last dipole of the beam line. The electrostatic deflector was controlled with a Standford Research Systems DG645 pulse generator which also synchronized the motion of the detection system (see below) and the data acquisition. The vacuum chamber (Fig.~\ref{fig:setup}) is split in two sections with independent pumping systems. In order to reduce contamination by $^6$He atoms, a stainless steel plate with a \O 35~mm hole separates the two sections. The beam was transported through two \O 6~mm collimators located at the entrance and exit of the first section of the chamber [Fig.~\ref{fig:setup} panel (a)]. During beam tuning, a movable Si detector was inserted at the entrance of the detection section to measure the incoming beam intensity.

The detection setup was designed for precision measurements of both, beta energy spectra
and half-lives of specific radioactive species. A $4\pi$ calorimetry coverage is achieved using two detectors. One of the detectors is fixed and aligned along the beam line whereas the other is mounted on a fast actuator controlled by the DG645 pulse generator. The $^6$He$^+$ beam is implanted on the front face of the fixed detector at a mean depth of about 130~nm as determined by TRIM, while the moving detector is located out from the beam [Fig.~\ref{fig:setup} (a)]. In this configuration, the beam passes across a third \O 4~mm collimator located a few mm from the fixed detector surface. 

The time sequence consists of an implantation interval of typically $T_{\rm imp} = 2.5$~s, followed by
a 1~s waiting interval during which the movable detector is brought in contact with the fixed detector, and finally a measuring interval of duration $T_{\rm acq}$. The movable detector remains in the measuring position [Fig.~\ref{fig:setup} (b)] during data taking before being lifted up to start over a new cycle.

This closed geometry with two detectors ensures the full collection of beta particles emitted by the implanted ions and prevents any partial energy loss due to backscattering. The duration of the data taking sequence was adjusted to be long enough such as to obtain a precise measurement of the ambient background. Here after, the fixed and movable detectors will be referred to as ``Det1'' and ``Det2'' respectively.
The transit times of Det2 were measured to be 0.8~s on the way down and 1.7~s on the way up. The loss of activity between the end of the implantation interval and the beginning of the measurement was thus limited to 50\%.
The total transit time of 2.5~s for a cycle duration of about 20~s does not lead to a significant loss of statistics. The motion of the detector was intensively tested for several days with 20~seconds duration cycles prior the experiment. These tests have shown no degradation of the detector signals. The fair contact between the two detector faces while in data taking position was also checked before and after the experiment by pinching 20~$\mu$m thickness sheets at different places of the contact surface.

\subsection{Beta particle detectors}
\begin{figure}[ht!]
\includegraphics[width=\columnwidth]{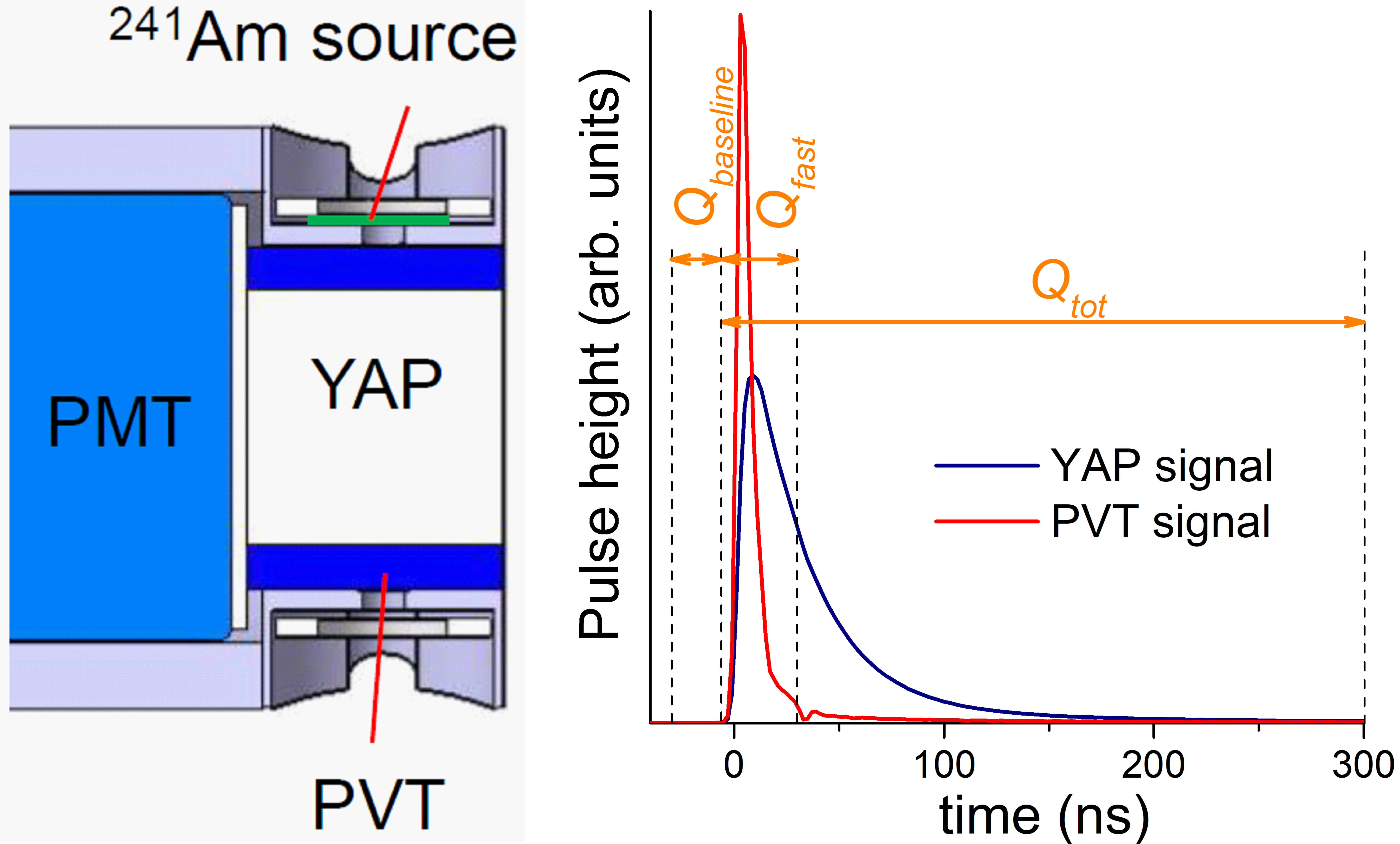}
\caption{Left panel: Close sectional view of a detector with the two scintillators coupled to a single PMT in a phoswich configuration and the location of the $^{241}$Am source. Right panel: Typical signals from the PVT and YAP along with the three charge integration intervals.}
\label{fig:detector}
\end{figure}

The two detectors, Det1 and Det2, are identical. Each of them is composed of a cylindrical \O$30 \times 30$~mm$^2$ YAlO$_3$ Ce doped inorganic scintillator (YAP)
surrounded by a EJ-204 plastic scintillator (PVT) with external diameter of
40~mm (Fig.~\ref{fig:detector}).
The scintillators are mounted in a phoswich configuration in which the YAP and the PVT are readout by a single R7723 photo-multiplier tube (PMT) from Hamamatsu. The aluminum detector housing serves in part as a shield and ensures also some pressure to produce a fair optical contact between the PMT, the optical coupling grease and the scintillators. Two circular hollows at the front of the housing allow the insertion of calibration sources.

A 5~kBq $^{241}$Am source was permanently mounted on each detector to provide a constant calibration reference using the 59.54~keV photons. These interact mostly in the YAP volume whereas the $\sim$5~MeV alpha particles are stopped in the PVT which also serves as a veto to reject background events. The detection of the alpha particles by the PVT is used to monitor PMT gain variations.
A $225~\mu$m
layer of Tyvek with a \O 5~mm hole in front of the $^{241}$Am source was wrapped around the PVT to improve the light collection. The faces of both detectors which get into contact are free of any reflector or dead layer. As a result, when scintillation light is generated in one of the detectors, about 80\% of the light is collected by its own PMT while the other 20\% is collected by the PMT of the other detector. 

The choice of YAP crystals resulted from a compromise between a fast response, an acceptable energy resolution as well as differential linearity over the largest possible energy range \cite{Mos98, Men98}. The YAP time decay constant of 27~ns is short enough to limit pile-up contributions but significantly longer than the PVT decay time of 2~ns. This leads to a clear discrimination between the PVT and YAP signals by pulse-shape analysis. Typical signals from the PVT and from the YAP scintillators are shown on the right panel of Fig.~\ref{fig:detector}. Moreover, the high photon detection efficiency of YAP allows the use of the photopeaks from gamma sources for off-line energy calibrations.
The nominal PMT polarization voltages used for Det1 and Det2 were $-1620$~V and $-1420$~V respectively. These values were chosen in order to ensure a linear response of the detectors over the full beta energy range (up to 3.5~MeV) and also to match their relative gain.

The response functions of the detectors were studied in detail prior to installing the detectors inside the vacuum chamber, using $^{241}$Am, $^{22}$Na, $^{137}$Cs and $^{60}$Co sources, as well as $\gamma$ rays from $^{208}$Tl. Small PMT gain variations between the measurements with different sources were corrected using the reference provided by the alpha particles interacting in the PVT.
Deviations from linearity were found to be smaller than 4~keV over the 60~keV--2.6~MeV range covered by the calibration sources. However, these deviations are most likely due to the fit of the photopeak which neglects the presence of the overlapping Compton or multi-Compton distributions. The energy resolution was 6.7\% at 662~keV and 5.5\% at 1332~keV.

Before and after the experiment, data were recorded again with the detectors under the experimental conditions and using $^{241}$Am, $^{22}$Na, $^{137}$Cs and $^{60}$Co sources to check the response function of the detectors.

\subsection{Data acquisition system}
The analog signals from the two PMTs and a logic pulse from the DG645 generator were directly sent to three input channels of the digital data acquisition system FASTER \cite{FASTER}. Each channel digitizes the signal at a rate of 500 MS/s with each sample converted over 12 bits. The triggers of the channels are independent based on individual thresholds. The digitized samples are processed
in real time by Field-Programmable Gate Arrays using predefined algorithms adapted to the measurements to be performed \cite{FASTER}. All samples are time-stamped with a 2 ns resolution allowing on-line and off-line correlations within user-defined time windows. For the two PMTs, the selected algorithm provided the charge integration of the signals within four different time windows relative to the trigger time. The first window was set between $-30$ and $-6$~ns to obtain the charge $Q_{\rm baseline}$ which provides an estimate of the baseline level (Fig.~\ref{fig:detector} right panel). The second and third windows were set between $-6$ and 30~ns and between $-6$ and 300~ns respectively. They enable to cover the PVT fast signals (charge $Q_{\rm fast}$) and the slower signals from the YAP (charge $Q_{\rm tot}$). The ratio $Q_{\rm fast}/Q_{\rm tot}$ serves for pulse-shape analysis and signal discrimination. A fourth window between $-30$~ns and 1~$\mu$s was also set to study after-pulses and pile-up effects. With these settings, the intrinsic dead-time of the acquisition system was 1030~ns. The logic pulse from the DG645 generator was used to provide a time reference for each measurement cycle.
The baseline of the three signals was continuously monitored for each channel and corrected for low-frequency variations (below 160 kHz) by the FASTER baseline restoration algorithm.
The trigger threshold of the two PMT channels was set to 5~mV, which corresponds to a minimum deposited energy of about 4~keV. The stability and accuracy of the internal clock of the FASTER system is smaller than 1~ppm and the associated systematic uncertainty on the half-life measurement is thus negligible relative to other corrections.

\section{DATA ANALYSIS}

In order to study systematic and background effects, five different experimental conditions were adopted resulting in five statistically independent data sets. The three parameters which were changed among the sets were the cycle duration, $T_{\rm acq}$, the initial activity and the bias voltages of the PMTs. A measurement of the ambient background with beam ON but without any implantation was also performed using a 0.4~mm thick aluminum disk fully covering the third collimator. For these runs, the incident beam intensity was adjusted so as to match the conditions of the short and long cycles runs at nominal PMT voltages, defining the conditions of sets (4) and (5) respectively.
The different sets and the corresponding number of cycles are summarized in Table~\ref{tab:conditions}. 

\begin{table}[!htb]
\caption{\label{tab:conditions} 
The five experimental conditions of the recorded data. $T_{\rm acq}$ is the duration of the measuring window; $\langle N_0 \rangle$ is the average number of detected $^6$He decays per cycle with an energy threshold of 300~keV; $\Delta V_{\rm PMT}$ indicates the difference in the PMT bias relative to the nominal voltage. The last column lists the number of recorded cycles.
}
\begin{ruledtabular}
\begin{tabular}{cccccc}
{Set}   & {Set name} & {$T_{\rm acq}$ (s)}  & {$\langle N_0 \rangle$}   & {$\Delta V_{\rm PMT}$ (V)} &  {Cycles} \\
\colrule
{(1)}       &{Short cycles}           &     {11}      &   {6130}         &   0      &   {3050} \\
{(2)}       &{Long cycles}            &     {26}      &  {20900}         &   0      &   {620} \\
{(3)}       &{Lower bias/Short}       &     {11}      &   {7140}         &   $-50$   &   {1930} \\
{(4)}       &{Shutter/Short}   &     {11}      &    $-$           &   0     &   {200} \\
{(5)}       &{Shutter/Long}    &     {26}      &    $-$           &   0      &   {115} \\
\end{tabular}
\end{ruledtabular}
\end{table}

For sets (1), (2) and (3), runs were devoted at the beginning of each
measurement to record signals waveforms which provided evidence of the presence of after-pulses in the range between 400~ns and 1~$\mu$s after the trigger. This prevented the use of the fourth charge integration window for the extraction of the $^6$He half-life.
 
\subsection{Baseline and gain corrections}
\label{sec:gainCorr}
A precise absolute energy calibration is not required for a precise half-life measurement. However, it is crucial to control baseline and gain variations within the decay cycle which could result in a change of the effective energy threshold imposed to the data.

\begin{figure}[ht!]
\includegraphics[width = 1.\columnwidth]{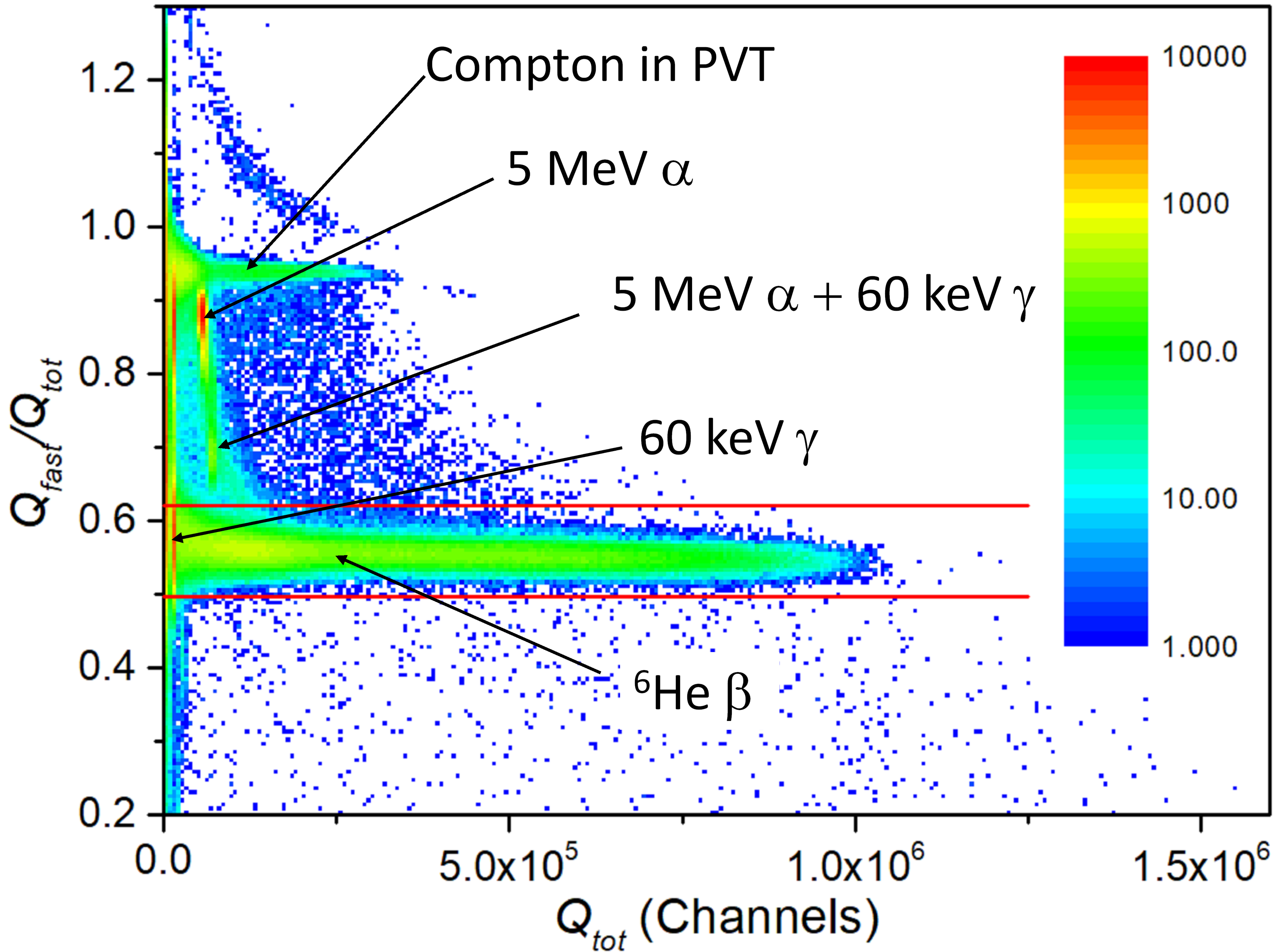}
\caption{2D-plot of the the ratio $Q_{\rm fast}/Q_{\rm tot}$ versus the total charge for signals from Det1. The contributions of the different particles are indicated. The horizontal red lines define the window of events due to an interaction in the YAP volume only. The plot was obtained for a one hour run under conditions of set (1).}
\label{fig:QfastQtot}
\end{figure}

The method described below is based on monitoring the time dependence of the 59.54~keV line from the $^{241}$Am source.
The calibrations of Det1 and Det2 were performed independently for each set of measurements. The method is illustrated here with data from set (1) and for Det1. First, a 2D-plot of the ratio $Q_{\rm fast}/Q_{\rm tot}$ versus $Q_{\rm tot}$ is built which enables particle identification through pulse-shape discrimination.
Events associated with scintillation light generated by 
the YAP only are then selected by setting a window over $Q_{\rm fast}/Q_{\rm tot}$ between 0.5 and 0.62 (Fig.~\ref{fig:QfastQtot}).

\begin{figure}[ht!]
\includegraphics[width = 1.\columnwidth]{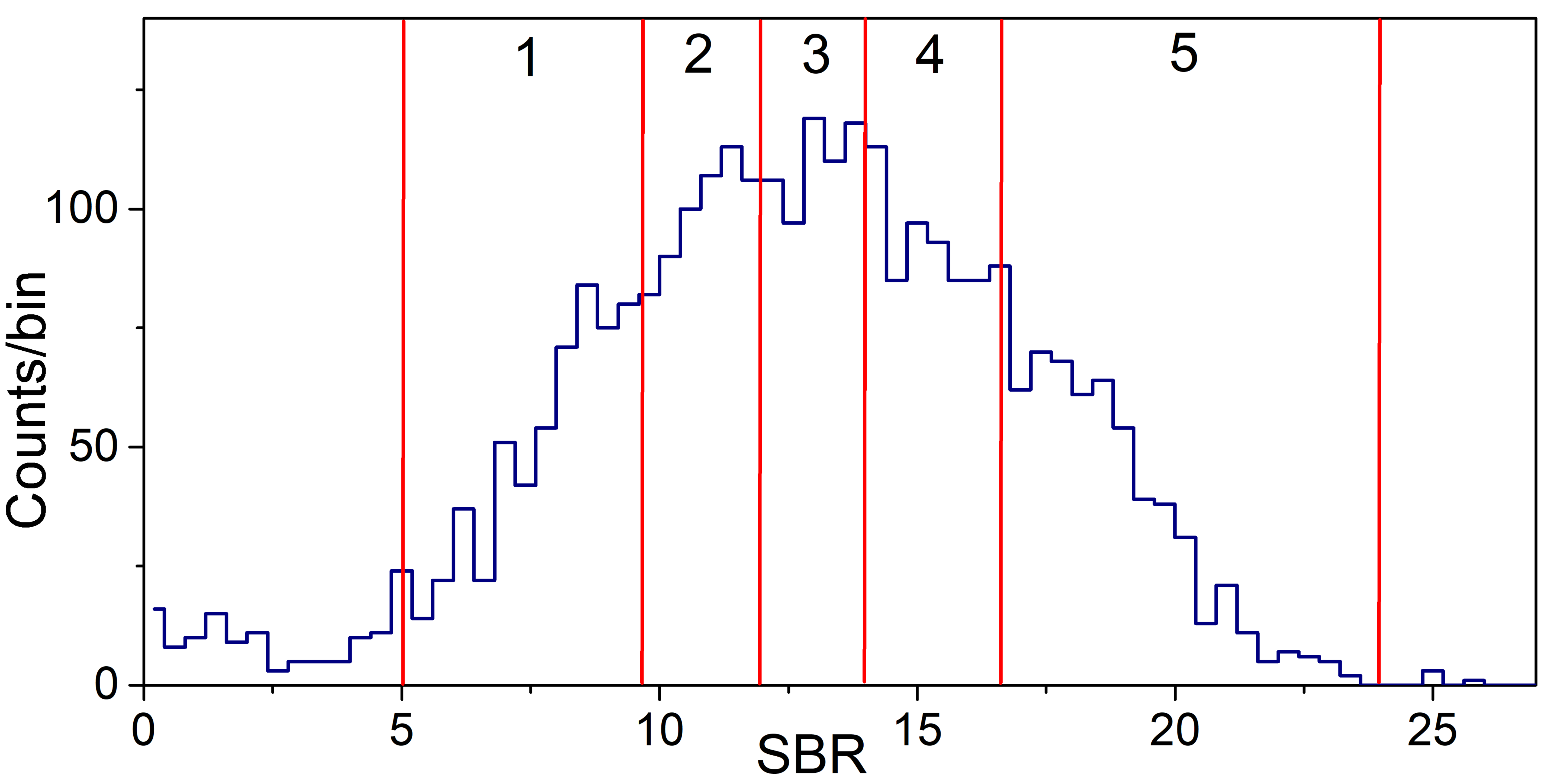}
\caption{\label{fig:SBR}{Distribution of SBR for set (1). The groups for different values of SBR, labelled from 1 to 5, are indicated by the vertical red lines.}}
\end{figure}

The cycles were then sorted as a function of the signal to background ratio (SBR) defined as the number of $^6$He decay events divided by the average constant background measured within a single cycle. The SBR was obtained for each cycle using a fit of the time distribution with an exponential decay plus a constant. The SBR distribution for set (1) is shown in Fig.~\ref{fig:SBR}. Five groups of SBR values containing comparable number of cycles were then defined. Cycles with SBR smaller than 5.0 were definitely discarded from the analysis as their very low statistics can lead to fitting convergence problems in the half-life extraction. The small cluster at SBR~$\sim25$ in Fig.~\ref{fig:SBR} was also excluded from the baseline and gain correction model because of its low statistics, but the corresponding cycles were included in the data used for the final half-life estimate.

\begin{figure}[ht!]
\includegraphics[width = 1.\columnwidth]{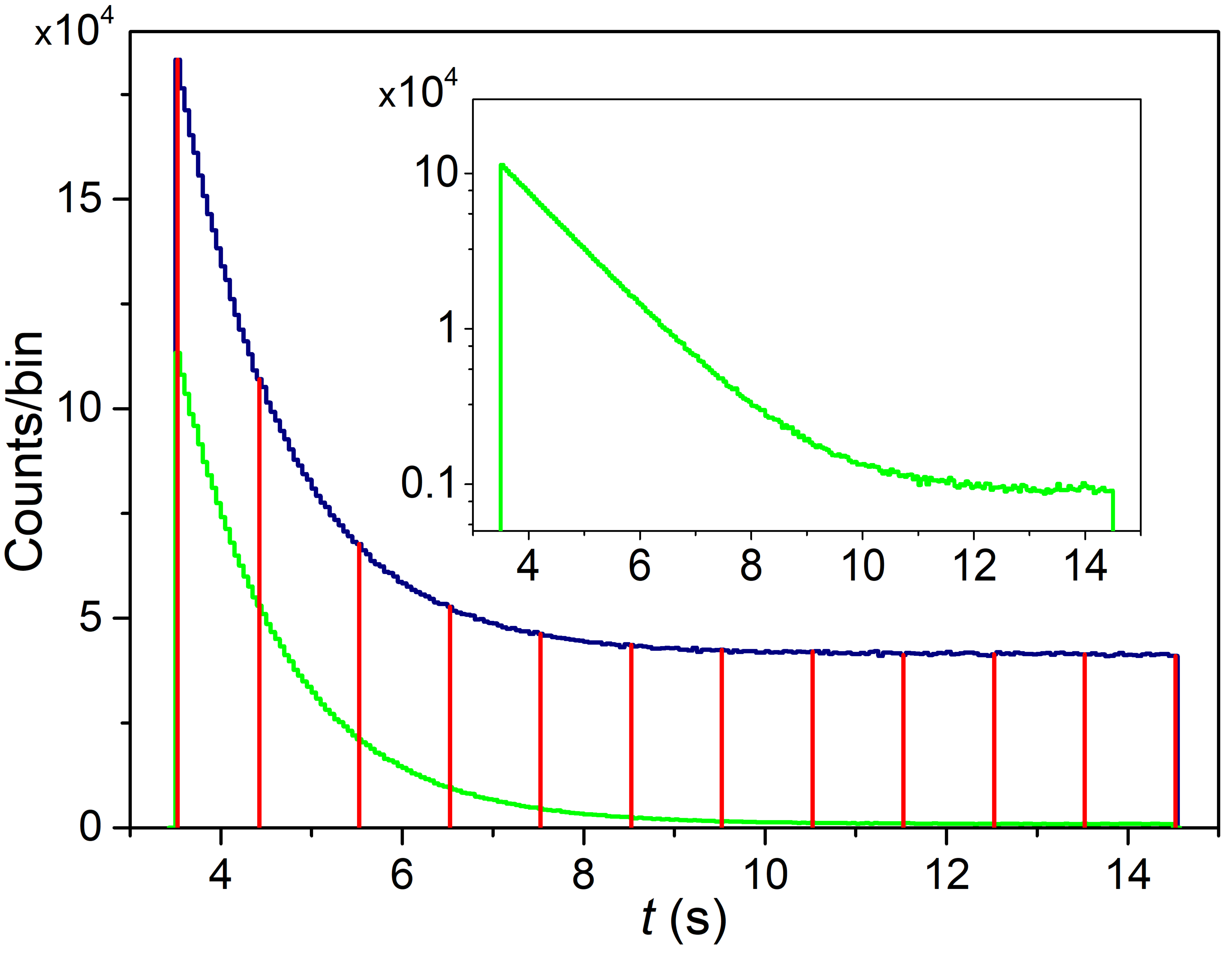}
\caption{\label{fig:decayshort}{Events collected during a one hour long run under conditions of set (1) as a function of the time within a cycle. The vertical red lines indicate the separation between the time windows selected for the gain variation correction. The beginning of the beam-on period is at $t=0$ and the measurement period starts at $t=3.5$~s. The blue histogram corresponds to the total number of events, including those from the $^{241}$Am source. The green histogram are those events remaining after the $Q_{\rm fast}/Q_{\rm tot}$ charge selection and an energy threshold of 100~keV. One bin corresponds to 50~ms.}}
\end{figure}

\begin{figure}[ht!]
\includegraphics[width = 1.\columnwidth]{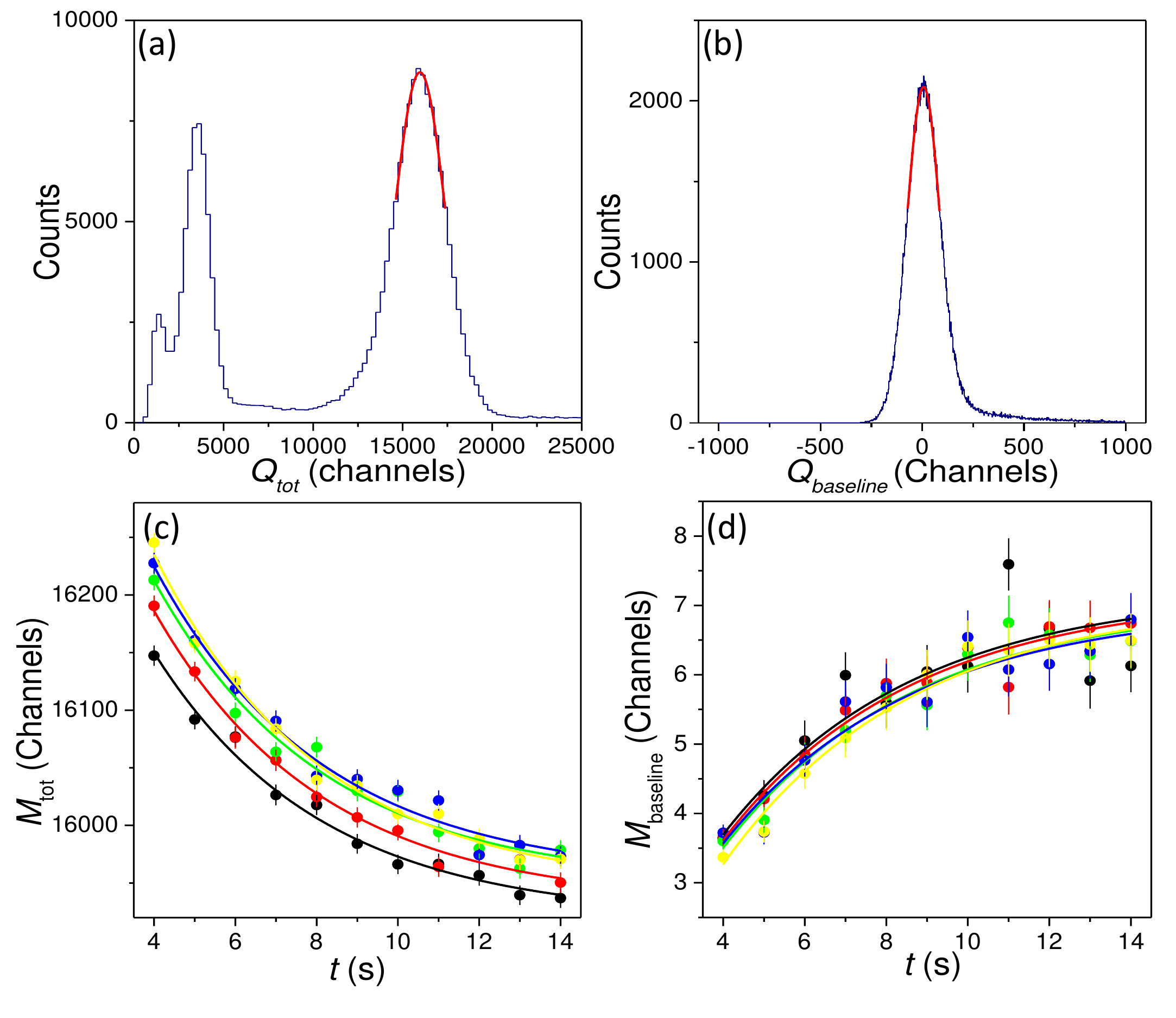}
\caption{(a) and (b): Close-up to the $Q_{\rm tot}$ and $Q_{\rm baseline}$ distributions associated to a SBR window and a time window. The red lines show the fits of the 59.54~keV photopeak (a) and of the baseline (b). (c) and (d): The data points show the mean values resulting from the fits as a function of time for the 59.54~keV photopeak (c) and the baseline (d). The black, red, green, blue and yellow colors correspond respectively to the SBR windows 1 to 5. The lines are the exponential fit functions obtained for each SBR set.}
\label{fig:gaussfit}
\end{figure}

For each group of cycles, events were then sorted according to their detection time within the cycle, using successive windows of 1~s duration (Fig.~\ref{fig:decayshort}).
For each set of events corresponding to a given time window and a given SBR window, the charge distribution $Q_{\rm tot}$ of the 59.54~keV photopeak and the $Q_{\rm baseline}$ distributions were fitted using Gaussian functions [Fig.~\ref{fig:gaussfit} (a) and (b)]. The mean values 
 ($M_{\rm tot}$ and $M_{\rm baseline}$) obtained from these fits are plotted in Fig.~\ref{fig:gaussfit} (c) and (d) as a function of time for the five different windows of SBR. A significant gain variation, close to 2\%, is observed in Fig.~\ref{fig:gaussfit} (c) between the beginning and the end of the measurement period. A baseline variation is also visible but has a smaller impact.
For each SBR window, the data of Fig.~\ref{fig:gaussfit} (c) and (d) were then fitted with functions of the form $P_0 + P_1 e^{-t/\tau_G}$ where $\tau_G$ is a common parameter for the five different SBR windows.

\begin{figure}[ht!]
\includegraphics[width = 1.\columnwidth]{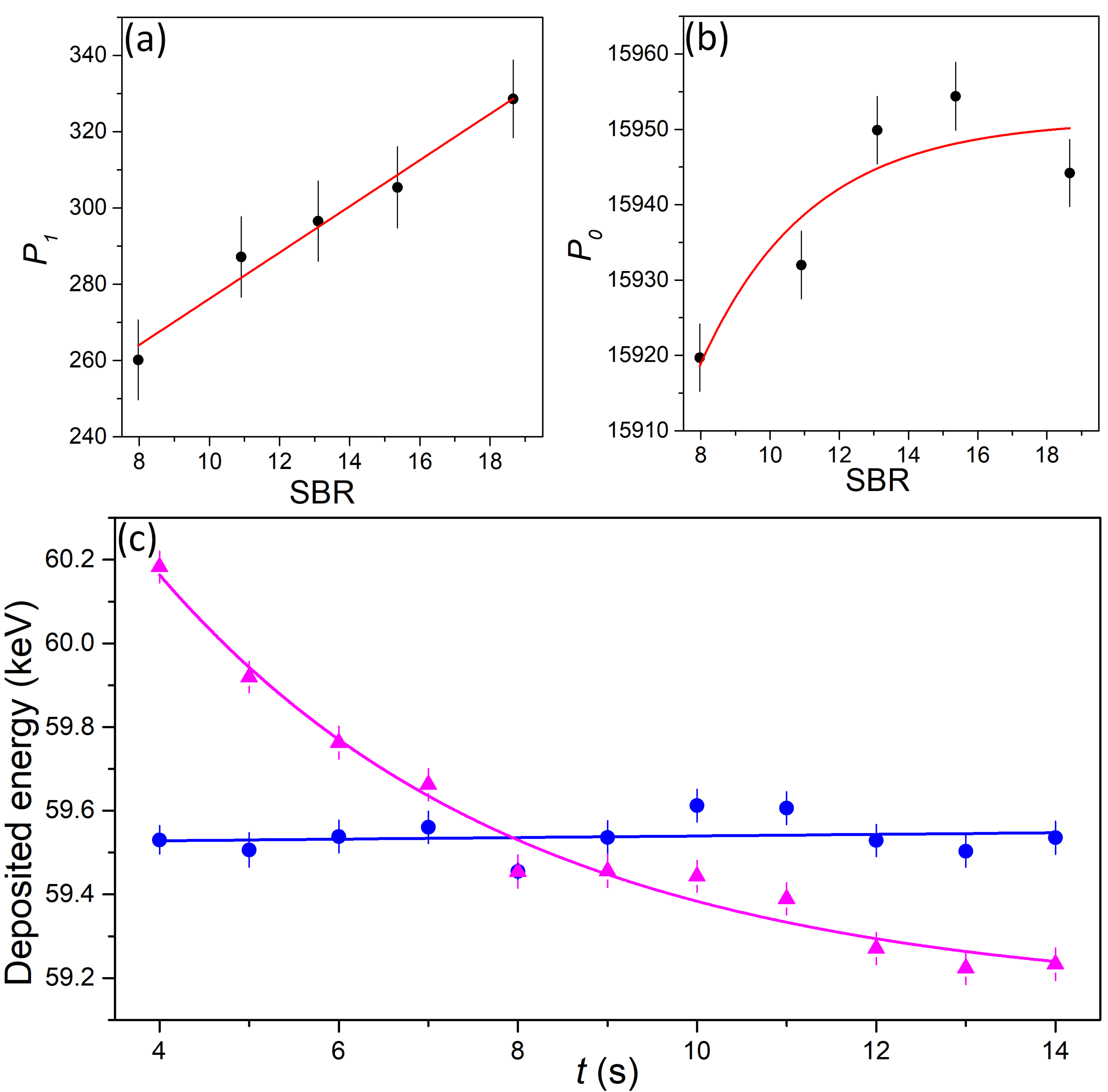}
\caption{\label{fig:SBRmodel}{(a) and (b): Parameters $P_1$ and $P_0$ for the gain correction model for the five values of SBR; (c) Results from fits of the 59.54~keV photopeak as a function of time without (magenta) and with (blue) the gain and baseline corrections.}}
\end{figure}

The gain and baseline correction models were completed by extracting parameters $P_0$ and $P_1$ from fits of the mean values of the charge distributions as a function of time for each SBR window. For the gain correction, the parameters are illustrated in Fig.~\ref{fig:SBRmodel} (a) and (b). The parameter $P_1$ is here the most relevant one since it causes a time dependence of the gain and of the baseline. It was found to depend linearly on the SBR value for both the gain and baseline distributions. For the gain correction model, $P_0$ was extracted using an exponential function whereas it was taken as constant for the baseline correction model.
%

Using these models, Det1 and Det2 were independently calibrated for each time window within a cycle and using the parameters associated with the SBR value of the cycle. Calibrated data from both detectors were then summed up for each event and scaled by a final calibration coefficient of 0.82, evaluated run by run, which accounts for the light cross-talk between the detectors for
photons of 60~keV. Fig.~\ref{fig:SBRmodel} (c) shows an example of the values from the fits of the 59.54~keV photopeak, as a function of time, with (blue) and without (magenta) applying the gain and baseline corrections. 

\subsection{Background sources}
\begin{figure}[ht!]
\includegraphics[width = 1.\columnwidth]{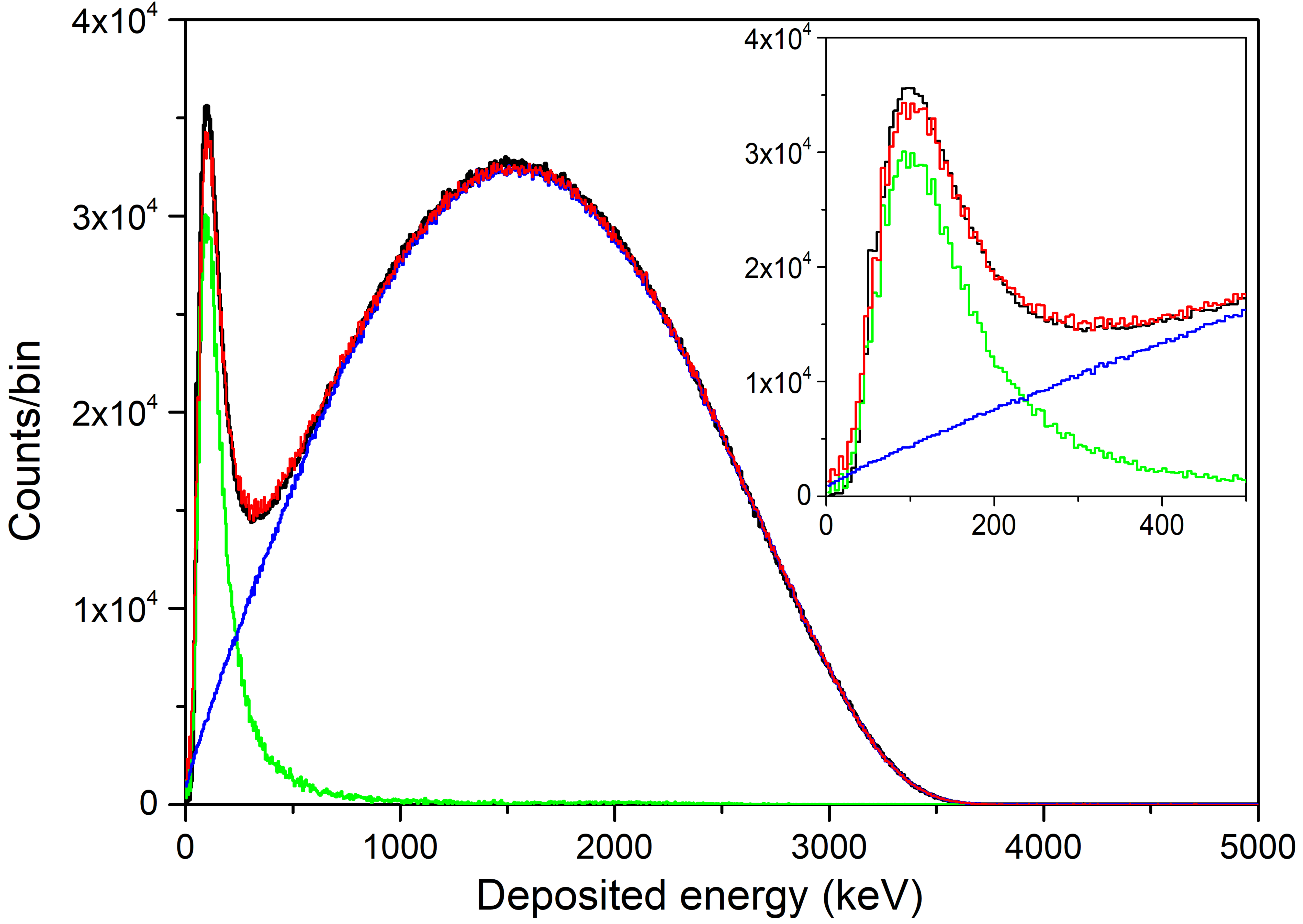}
\caption{\label{fig:DecaySpectrum}{Energy spectra for set (1) (black histogram) and (4) (green histogram) after background subtraction. The blue histogram is the theoretical beta energy spectrum including the convolution with the detector response function. The red histogram is the fit to the black spectrum with a combination from the blue and green spectra. The green and blue distributions are normalized using the parameters from the fit.}}
\end{figure}

The background rate was assumed to be constant during a cycle. The measuring cycle was divided in two halves and the energy spectrum corresponding to decay events was then obtained by subtracting the spectrum recorded during the second half from the spectrum recorded during the first half. The resulting energy spectra obtained for sets (1) and (4) are shown in Fig.~\ref{fig:DecaySpectrum}.
The black histogram corresponds to set (1) and shows two contributions. The first is the beta energy spectrum extending up to 3.5~MeV and the second is a low energy distribution with a peak at about 0.1~MeV. The decay time of this distribution is consistent with the $^6$He half-life and thus the contribution
was attributed to Bremsstrahlung radiation of electrons from $^6$He$^+$ decay, for ions implanted in the third collimator. This was confirmed by the analysis of data from set (4), shown by the green histogram in Fig.~\ref{fig:DecaySpectrum}. We recall that the data for this set was obtained with a shutter located inside the third collimator in order to prevent any implantation on the detector. The theoretical beta energy spectrum convoluted with the detector response function, is shown by the blue histogram. The sum of the blue and green distributions is displayed by the red histogram which fairly reproduces the data from set (1) (insert in Fig.~\ref{fig:DecaySpectrum}). This Bremsstrahlung background does not affect the measurement since it decays with the same half-life. Furthermore, by setting an energy threshold above 0.6~MeV, its contribution represents 0.5\% of the number of events associated with the beta particle detection.

\begin{figure}[ht!]
\includegraphics[width = 1.\columnwidth]{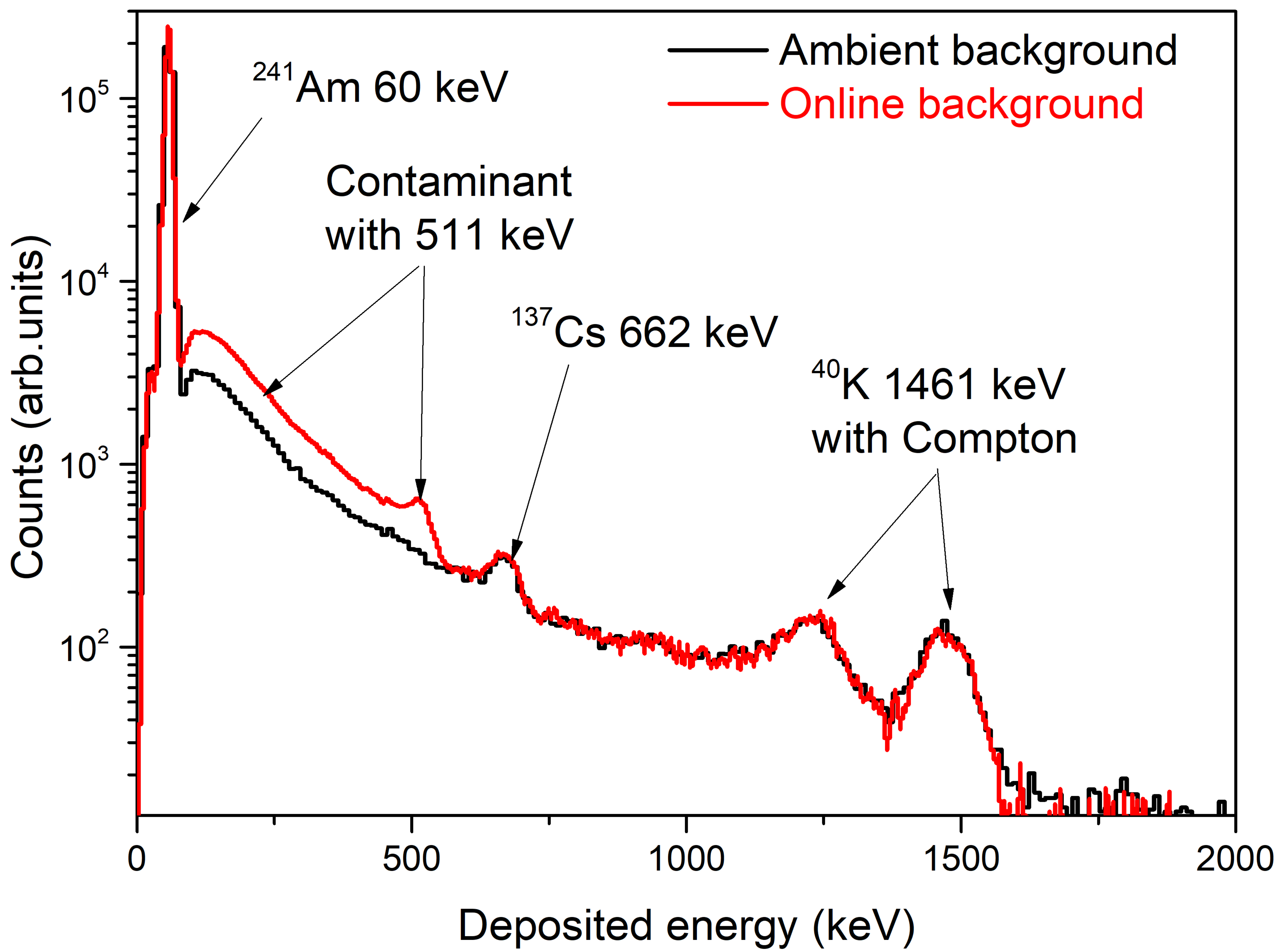}
\caption{Energy spectra for ambient background (black histogram) and for constant or long half-life background recorded on-line for one run of data set (1) (red histogram). The black histogram was normalized to the red one by using the 59.54~keV peak from $^{241}$Am.}
\label{fig:BKGSpectrum}
\end{figure}

Conversely, the selection of data in the second half of the measurement cycle suppresses the contribution due to $^6$He decay. In order to fully subtract  this contribution, a proper normalization factor was used on the spectrum measured during the first half of the cycle. This normalization factor was determined assuming that events with an energy deposit larger than 2~MeV were solely due to $^6$He decay. The resulting energy spectrum is shown in Fig.~\ref{fig:BKGSpectrum} (red histogram) and corresponds to a constant or long half-life background compared to $^6$He. The peak at 59.54~keV is from the monitoring $^{241}$Am source and the photo-peaks at 662~keV and 1461~keV are from a $^{137}$Cs source of a dose rate meter located in the experimental area and from ambient $^{40}$K respectively. 
Compared to the ambient background spectrum recorded before and after the on-line experiment (black histogram), an excess of events is observed below 0.6~MeV, with a peak due to 511~keV photons. 
This contaminant was not clearly identified and its half-life was found too long to be measured with the cycle length chosen for the experiment. A lower limit of 330~s (95\% CL) could however be extracted from this background analysis. The average detection rate of this contaminant contribution was found to be $\sim$14 cps and it can be fully suppressed by applying an energy threshold of 0.6~MeV. The ambient background contribution was constant during the experiment.
%

\subsection{Dead-time correction}
To study systematic effects due to dead time, a common recipe \cite{Kne12b} consists in using analog parallel channels with different non-extensible dead times in the data acquisition. With precisely time-stamped events recorded by the digital system, the data were first filtered off-line by imposing several dead times $\tau_{DT}$ ranging from 1.03 to 7~$\mu$s. In order to study further possible bias due to background contributions or imperfections of the gain and baseline corrections, the data were additionally filtered using energy thresholds ranging from 0.1 to 1.2~MeV.

The probability to miss a decay event at time $t$ within a cycle because of dead time can be expressed as $\tau_{DT}\cdot r_T(t)_i$, where $\tau_{DT}$ is the non-extensible dead time and $r_T(t)_i$ is the total instantaneous rate of events that could potentially trigger the data acquisition system during cycle $i$. This rate is given by
\begin{equation}
r_T(t)_i = r_{0i} e^{ -t/\tau_{i}} + r_{bi}~,
\label{eq:rtot}
\end{equation}
with $r_{0i}$ the total initial decay rate and $r_{bi}$ the total rate of background events. First, the parameters $r_{0i}$, $\tau_{i}$ and $r_{bi}$ which enter Eq.(\ref{eq:rtot}) were obtained by fitting the rate of detected events for each cycle, $r_D(t)_i$, without threshold condition, and using the correction accounting for losses due to dead time
\begin{equation}
r_D(t)_i = \frac{r_T(t)_i}{1 + \tau_{DT}\cdot r_T(t)_i}~.
\label{eq:rdet}
\end{equation}
If only the dead-time effect was present,
for a given dead time $\tau_{DT}$, the occurrence of an event at time $t$
should be weighted by a coefficient
\begin{equation}
w(t)_i = 1 + \tau_{DT} \cdot r_T(t)_i~.
\label{eq:weight}
\end{equation}
However, pile-up effects have also to be considered and can be included
in the weighting coefficient.
%
\subsection{Pile-up effects}
An off-line energy threshold condition, labelled $j$, was
also applied to the data. With this condition, the rate of selected events is then given by 
\begin{equation}
r_D(t)_{ij} = r_D(t)_i \cdot P_E(t)_j~,
\label{eq:rdetj}
\end{equation}
where $P_E(t)_j$ is the probability for the measured deposited energy to be above the energy threshold $E_j$.
To first order, $P_E(t)_j$ should be independent of time but the occurrence of pile-up events during the charge integration window can lead to a small time dependent correction. When a pile-up occurs within the window, the recorded charge and its associated energy will naturally be larger than for each of the events producing the pile-up. The recorded event has then a larger probability to result in an energy above the threshold.
The pile-up probability is proportional to the instantaneous rate in the decay cycle. This leads to an excess of events with an energy above threshold for high rates as compared to low rates and would result in an underestimation of the half-life if not properly corrected.

The instantaneous energy-independent probability for a decay event to pile-up with another decay event within a cycle $i$ is given by
\begin{equation}
P_{\rm pu}(t)_i = r_{0i}e^{-t/\tau_{i}} \Delta T~,
\label{eq:Ppu}
\end{equation}
where $r_{0i}$ and $\tau_{i}$ are the parameters determined from the fit of Eq.(\ref{eq:rtot}) and $\Delta T = 300$~ns is the duration of the charge integration window.

The excess of detected events due to pile-up can then be expressed as 
\begin{equation}
r_E(t)_{ij} \approx P_{\rm pu}(t)_i \cdot (d_j - s_j)
r_{0i}e^{-t/\tau_{i}}~,
\label{eq:r_excess}
\end{equation}
where $s_j$ and $d_j$ are respectively the time independent probabilities for an energy conversion to be above threshold for single events and for pile-up events involving two signals. The contribution to the half-life of pile-up events involving three or more signals was estimated to be smaller than $10^{-7}$~s and hence negligible. On the other hand, the expected rate without pile-up contribution would be
\begin{equation}
r_R(t)_{ij} = s_j\cdot r_{0i}e^{-t/\tau_{i}} + b_j\cdot r_{bi}~,
\label{eq:r_real}
\end{equation}
where $b_j$ is the time independent probability to be above threshold for background events.

The probabilities $s_j$ and $b_j$ to get a signal above threshold $E_j$ for single events were calculated from the deposited energy distribution of $^6$He decay events and background events, respectively (Figs.~\ref{fig:DecaySpectrum} and \ref{fig:BKGSpectrum}). 
The expected energy distribution for $^6$He pile-up events was deduced using the auto-convolution of the energy distribution obtained for $^6$He single events. The mean probability, $d_j$, for a pile-up event within 300~ns to be above threshold was then determined. The fraction of the charge of the second signal that is lost when overlapping with the end of the integration window was accounted for in this calculation. 
Note that the effect of pile-up events involving the constant ambient background and the contaminant contribution discussed in section III.B were neglected in Eq.(\ref{eq:r_excess}) since the associated correction was estimated to be below 10$^{-6}$~s. 

Finally, the relative rate excess of detected events as a function of time can then be approximated by
\begin{equation}
\alpha_E(t)_{ij} = \frac{r_E(t)_{ij}}{r_R(t)_{ij}}~,
\label{eq:r_rel_excess}
\end{equation}
and the weighting coefficient which accounts for both, dead-time and pile-up effects becomes
\begin{equation}
w(t)_{ij} = \frac{1 + \tau_{DT} \cdot r_T(t)_i}{1 + \alpha_E(t)_{ij}}~.
\label{eq:weight2}
\end{equation}
For simplicity in the discussion above, a given value of the dead time, $\tau_{DT}$, was taken in all expressions. Changing this value corresponds to an additional condition,
labelled $k$ here below, so that the weights in Eq.(\ref{eq:weight2})
become $w(t)_{ijk}$.
\subsection{Fit procedure}
\label{sec:fit}
After applying a given dead time and energy threshold,
the weighted events from all cycles within a measurement set were summed-up and binned.
The weighted number of counts in a  bin is then $n(t)_{jk} = \sum_{i} {w(t)_{ijk}}$
and the variance is $\sigma^2_{jk} = \sum_{i} {w^2(t)_{ijk}}$,
where the sums run over the cycles and also over all events within each cycle.
The corrected data were finally fitted assuming a constant background. For a given dead time and energy threshold, the fit function can be expressed as
\begin{equation}
f(t)_{jk} = A_{jk}\cdot e^{ -t/\tau_{jk}} + B_{jk}~,
\label{eq:fitFct}
\end{equation}
where $A_{jk}$ is the initial number of decay counts, $\tau_{jk}$ is the estimate of
the decay lifetime and $B_{jk}$ is the constant background level.

To summarize, the fit procedure for each set involves three steps. First, the
rate of detected events without energy condition was fitted for each cycle with Eq.(\ref{eq:rdet}) with $r_{0i}$, $\tau_{i}$ and $r_{bi}$ as free parameters, to be used in the dead-time and pile-up corrections. Next, the dead time and energy threshold corrections were estimated and the
occurrence of each event was weighted using Eq.(\ref{eq:weight2}). Finally, weighted events from a measurement set were summed-up and the resulting binned distribution was fitted with Eq.(\ref{eq:fitFct}), with $A_{jk}$, $\tau_{jk}$ and $B_{jk}$ as free parameters.
Following this procedure, all the data from a measurement set results in a single
histogram.
For all fits, the parameter estimates were obtained using the log-likelihood function. The nominal bin size adopted in the histograms was 50~ms.

\begin{figure}[ht!]
\includegraphics[width = 1.\columnwidth]{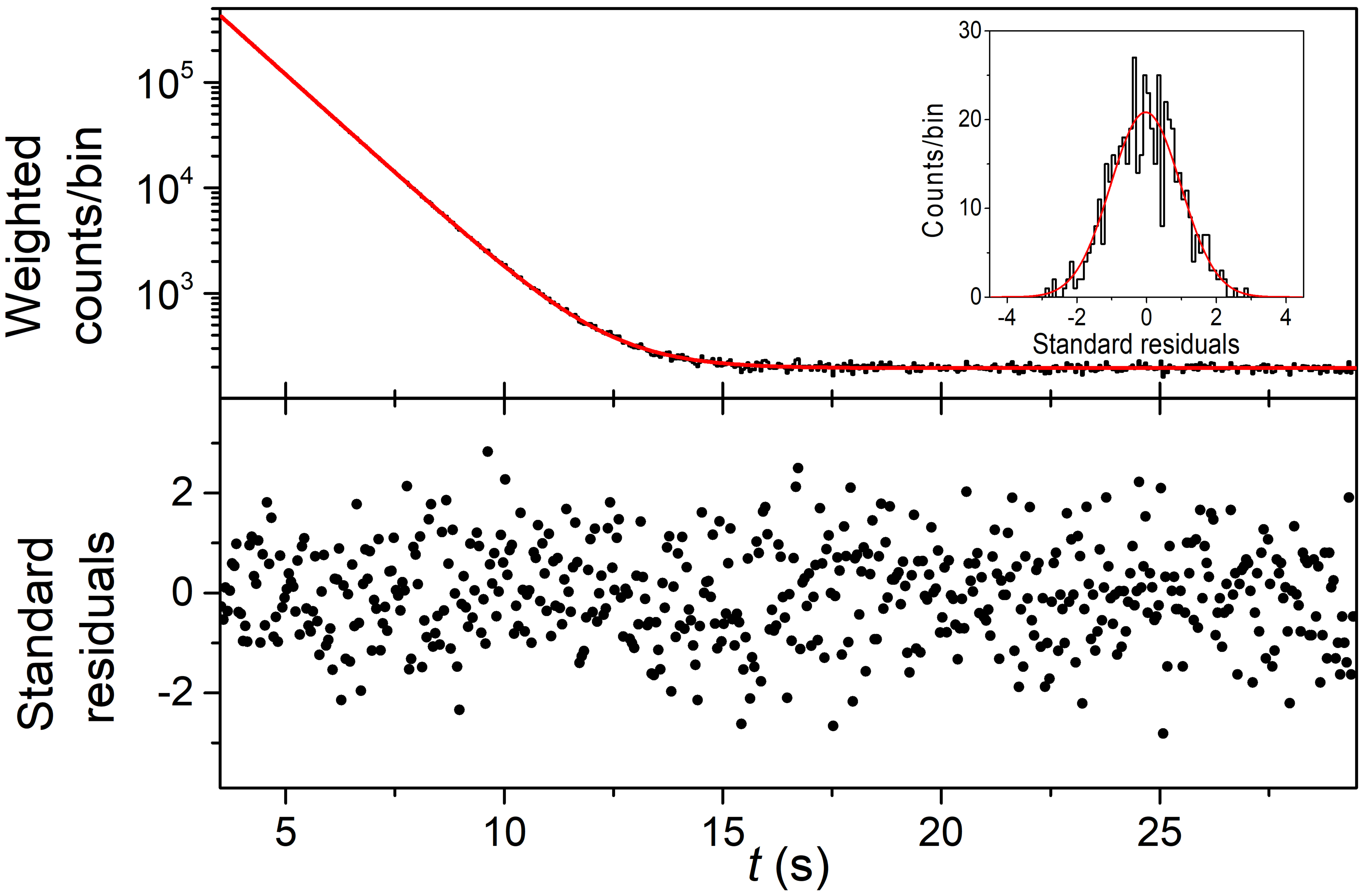}
\caption{Upper panel: experimental decay spectrum from set (2) for a non-extensible dead time of 3~$\mu$s and an energy threshold of 600~keV (black) along with its fit function (red). Lower panel: standard residuals. The standard residuals distribution and its fit by a Gaussian are shown in the insert of the upper panel.}
\label{fig:decay_fit}
\end{figure}

\begin{figure}[ht!]
\includegraphics[width = 1.\columnwidth]{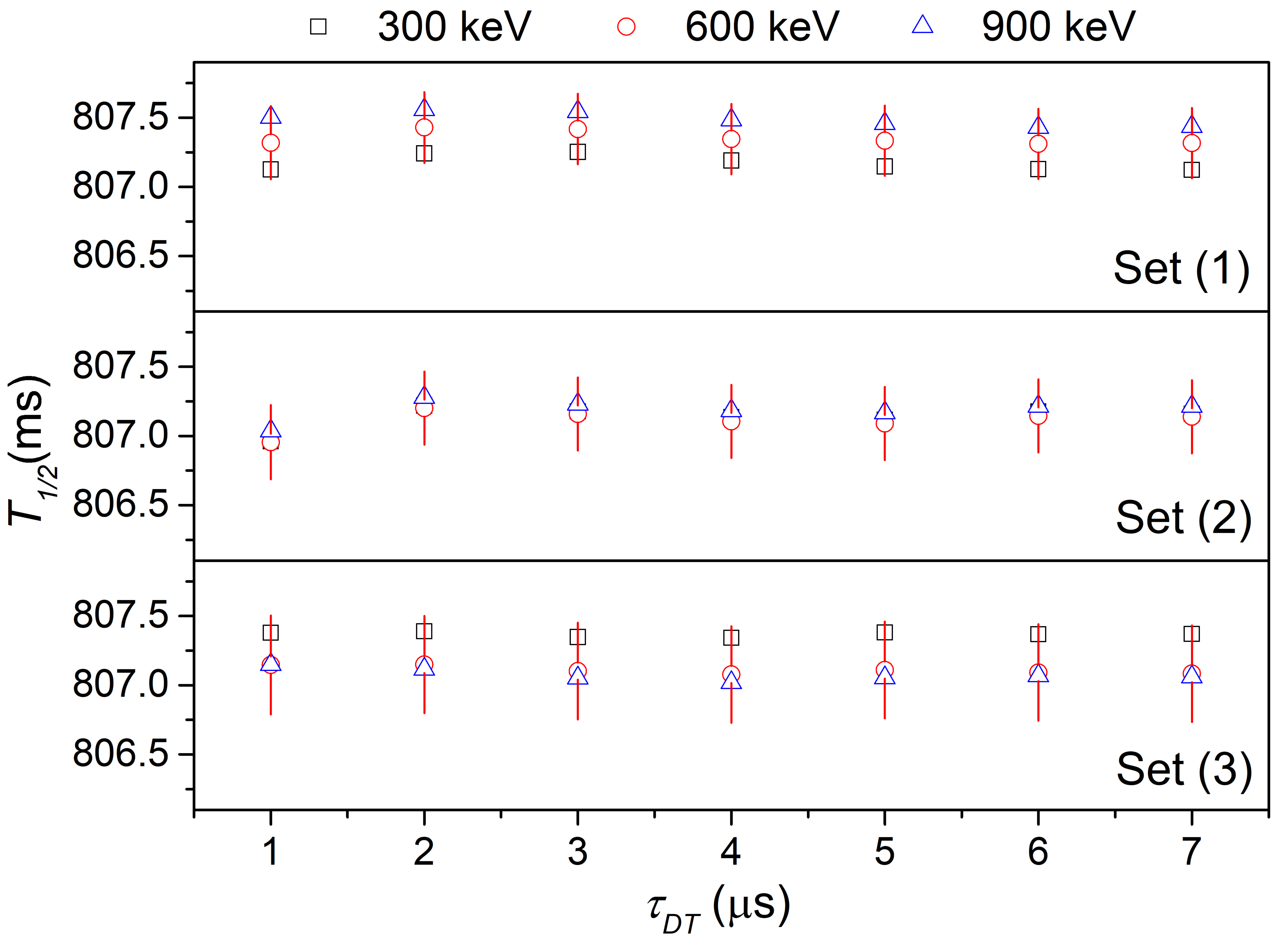}
\caption{Mean values of the half-life estimates obtained for set (1) (upper panel), (2) (middle panel) and (3) (lower panel) as a function of the non-extensible dead time. The black squares, red dots and blue triangles correspond respectively to off-line energy thresholds of 300, 600 and 900~keV. The statistical error bars are only shown for the threshold of 600~keV.}
\label{fig:T12_vs_DT}
\end{figure}

Figure~\ref{fig:decay_fit} shows the fit of data set (2) with a dead time of 3~$\mu$s and an energy threshold of 600~keV. The fit of the standard residuals distribution is consistent with a normal distribution with $\mu = -0.027(37)$ and $\sigma = 1.01(4)$.
%
In Table~\ref{tab:conditions}, the sets from which the half-life can be extracted are (1), (2) and (3). The resulting half-life values are shown in Fig.~\ref{fig:T12_vs_DT} as a function of the imposed dead time and for three energy thresholds. The values show no significant effect due to dead time except for set (2) and $\tau_{DT} = 1~\mu$s, which is slightly lower than the others. This discrepancy may be due to the contribution of after-pulses in the $1~\mu$s - $2~\mu$s range. In the following, a conservative value $\tau_{DT} = 3~\mu$s was adopted. The systematic uncertainty on the dead-time correction was obtained from the 2~ns accuracy on the imposed dead time and is smaller than 10$^{-5}$~s for all sets. 
The results from the fits of the histograms for the three sets are summarized in Table~\ref{tab:T12Values}. The central values include the systematic corrections due to dead time and pile-up but the uncertainties
are only statistical.

\begin{table}[!htb]
\caption{Values of the $^6$He half-life obtained from the fits of histograms for the
three data sets along with their associated $p$-value.}
\label{tab:T12Values}
\begin{ruledtabular}
\begin{tabular}{lccc}
               & Set (1) 	 & Set (2)     & Set (3)\\
\colrule
$T_{1/2}$ [ms]          &  807.42(25)     &     807.16(26)    &   807.10(35)\\
$p$-value            &  0.70       &     0.83      &   0.25\\
\end{tabular}
\end{ruledtabular}
\end{table}

\subsection{Effect of baseline and gain corrections}
\begin{figure}[ht!]
\includegraphics[width = 1.\columnwidth]{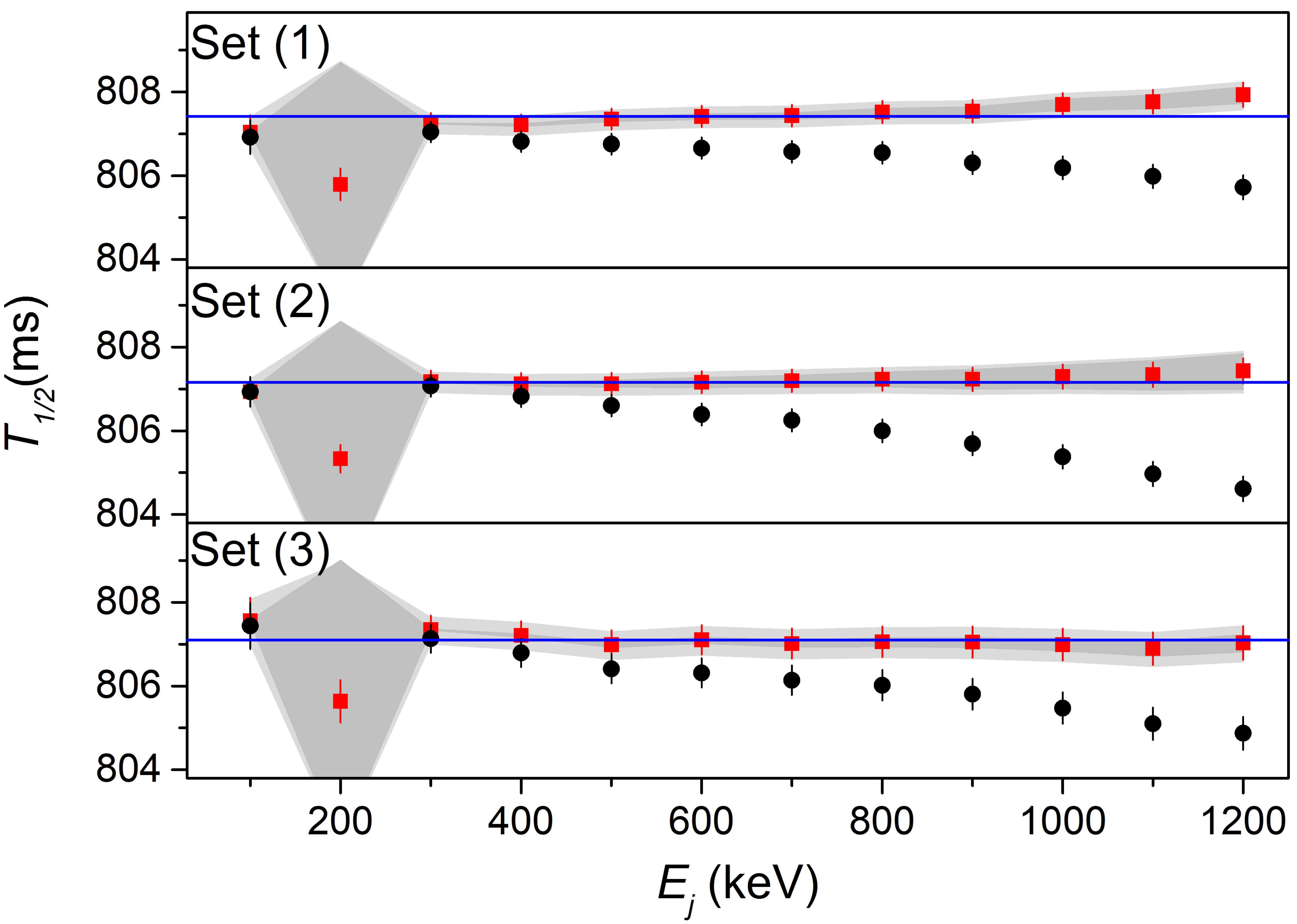}
\caption{Mean values of the half-life obtained for set (1) (upper panel), (2) (middle panel) and (3) (lower panel) as a function of the energy threshold. The black circles (resp. red squares) are the values obtained without (resp. with) the gain correction. The error bars are statistical. The horizontal blue line shows the value with correction for a threshold of 600~keV. The gray area indicates the systematic uncertainty associated to the gain correction and the light gray one the combined systematic and statistic uncertainty (see text for details).}
\label{fig:T12_vs_Eth}
\end{figure}

For illustration, Fig.~\ref{fig:T12_vs_Eth} shows, for the three sets, the evolution of the mean values of the half-life as a function of the energy threshold, before (black circles) and after (red squares) applying the gain correction described in Sec.~\ref{sec:gainCorr}. 
Whereas the non-corrected data display a trend which increases with the threshold between 300 and 1200~keV, the corrected data lead to mean values which are independent of the threshold.
The results obtained without gain correction for an energy threshold of 200~keV are out of the vertical range of the figure, at about 20~ms higher than with gain correction. This is due to the strong negative slope of the energy spectrum at 200~keV, right from the Bremsstrahlung peak shown in Fig.~\ref{fig:DecaySpectrum}. The systematic uncertainty associated to the gain correction procedure is shown by the gray areas. It was estimated by accounting for the uncertainties on the parameters $P_0$ and $P_1$ of the model and taking values which maximize or minimize the gain correction amplitude. As expected, this uncertainty also increases with the energy threshold between 300 and 1200~keV and is very large for a threshold of 200~keV. The distribution of the corrected values as a function of the threshold is consistent with a constant when accounting for both statistical and systematic uncertainties. In order to minimize the uncertainty on the gain correction, one should normally favor a threshold in the range 300-400~keV. However, because of the presence of background contributing up to 550~keV whose half-life is unknown (Fig.~\ref{fig:BKGSpectrum}) a threshold of 600~keV was finally chosen. The effect of the baseline correction was studied in a similar way and was found to be one order of magnitude smaller than for the gain correction. The $\chi^2$ and residuals distributions obtained for all sets where also studied with and without gain corrections. In both cases, they showed no deviation from the expected statistical fluctuations, which indicates that analyses of $\chi^2$ or $p$-values do not provide a proper diagnostic to detect such systematic effects.

\subsection{Diffusion of $^6$He}
The possible rapid diffusion of $^6$He atoms out of the detector bulk was considered. Diffusion coefficients specific to helium implanted in a YAP crystal could not be found in the literature. However, experimental data are available for a number of mineral compounds \cite{Tro14}. Diffusion coefficients at room temperature are all smaller than 10$^{-26}$~m$^2$s$^{-1}$. Using Fick's second law, an implantation depth of 100~nm leads to effusion time constants larger than 10$^{11}$s. Such time constants are far too large to have any significant effect on the $^6$He half-life measurement.
\begin{figure}[ht!]
\includegraphics[width = 1.\columnwidth]{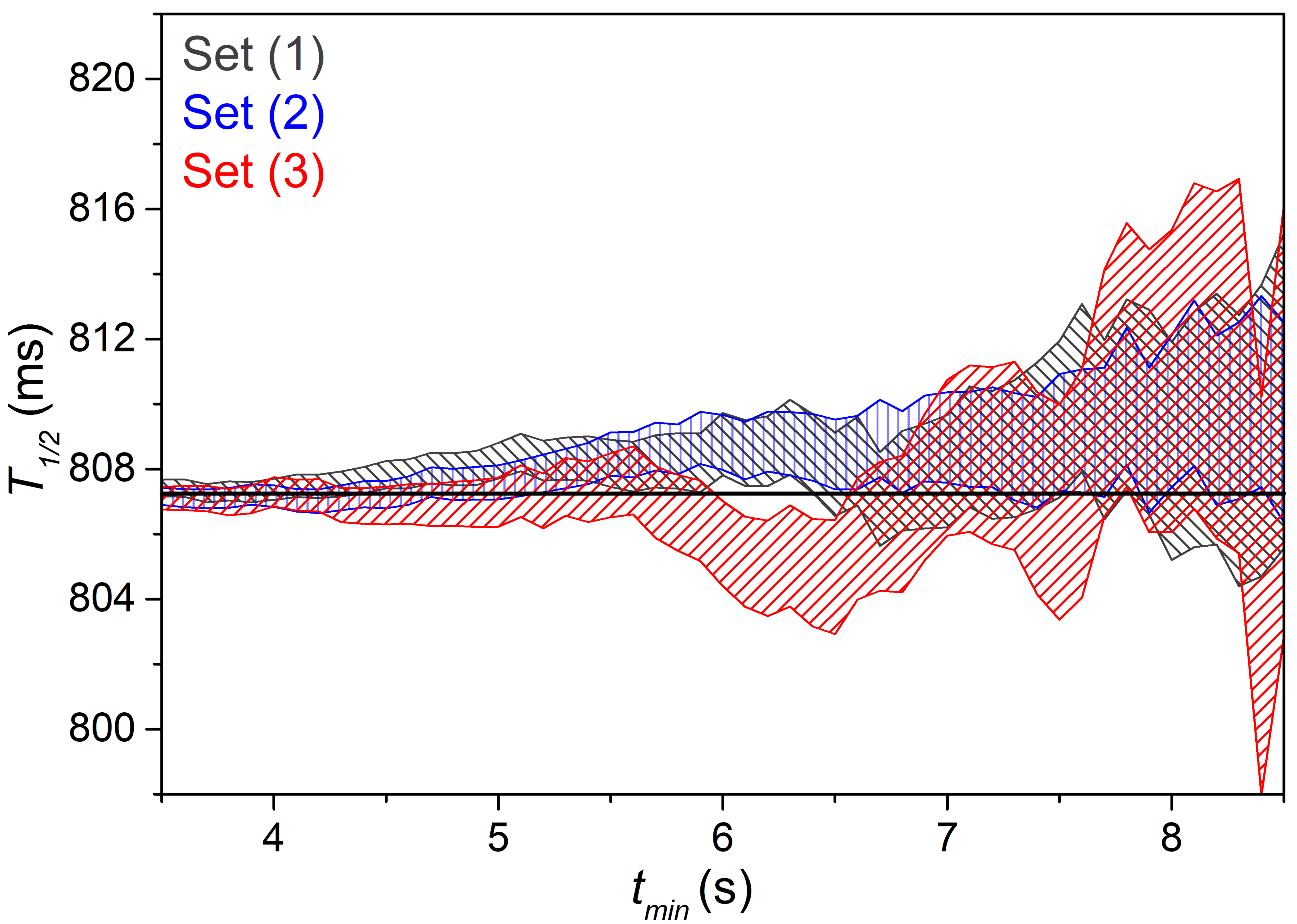}
\caption{\label{fig:Tfit_min}{Half-life estimate obtained for the measurements sets (1), (2) and (3) (in gray, blue and red, respectively) as a function of the lower bound of the fitting range, $t_{min}$. The colored area indicate the range covered by the statistical uncertainties given at one standard deviation. The horizontal black line is the central value of the final result.}}
\end{figure}
\subsection{Additional cross-checks}
The bin width of the histograms was changed from the nominal 50~ms to 20 and 100~ms to check the sensitivity to the binning. The resulting values of the half-life were found to be the same within 1$\times 10^{-5}$~s.

The dependence of the half-life on the lower bound of the fitting range, $t_{min}$ was also investigated (Fig.\ref{fig:Tfit_min}).
The analysis of $p$-values spanning more than 6 half-lives indicates that the observed
variations are fully consistent with statistical fluctuations within the three sets.

In the analysis procedure presented above, the data was corrected and then fitted with Eq.(\ref{eq:fitFct}).
An additional analysis was also carried out without applying any dead time and pile-up corrections to the data.
Instead, for a given condition in dead time and energy threshold, the data in each individual cycle $i$ was fitted with a function accounting for both dead time and pile-up effects
\begin{equation}
f_{ij}(t) = \frac{(A_{ij}\cdot e^{ -t/\tau_{ij}} + B_{ij}) [1 + \alpha_E(t)_{ij} ]}{1 + \tau_{DT}\cdot r_{T}(t)_i}~,
\label{eq:rij}
\end{equation}
where $r_{T}(t)_i$ and $\alpha_E(t)_{ij}$ are the functions in Eqs.(\ref{eq:rtot}) and (\ref{eq:r_rel_excess}) determined for each cycle and selection and where $A_{ij}$, $\tau_{ij}$ and $B_{ij}$ are the free parameters of the fit.
For each measurement set, the lifetime estimates, $\tau_{ij}$, obtained from the fits were averaged to obtain the half-life $\tau_{j}$ for that condition. This averaging required a closer look because the low statistics present in many cycles was found to induce a correlation
between the half-life and the absolute statistical uncertainty.

In order to test the averaging procedures, Monte Carlo simulations were performed using typical experimental decay and background rates and a higher number of decay cycles. The simulations showed that when using the relative statistical uncertainty instead of the absolute uncertainty as the weighting factor, the bias due to low statistics in the fitted data was strongly reduced. Nevertheless, the weighted average was found to be overestimated by 9$\times10^{-5}$~s for the conditions of sets (1) and (3) and by 1.5$\times10^{-5}$~s for the condition of set (2).
After accounting for the bias due to fits with low statistics, the final experimental values obtained when fitting independently each cycle were found to be the same within 1$\times 10^{-5}$~s
to those obtained using the method described in Sec.~\ref{sec:fit}.

The complete analysis was performed again while using the result for the half-life obtained in Eq.(\ref{eq:T12})
to fix the value of $\tau_{i}$ in Eqs.(\ref{eq:rtot}) and (\ref{eq:rdet}). The results were again found to be the same within 1$\times 10^{-5}$~s.

\section{RESULT}

\begin{figure}[!hbt]
\includegraphics[width = 1.\columnwidth]{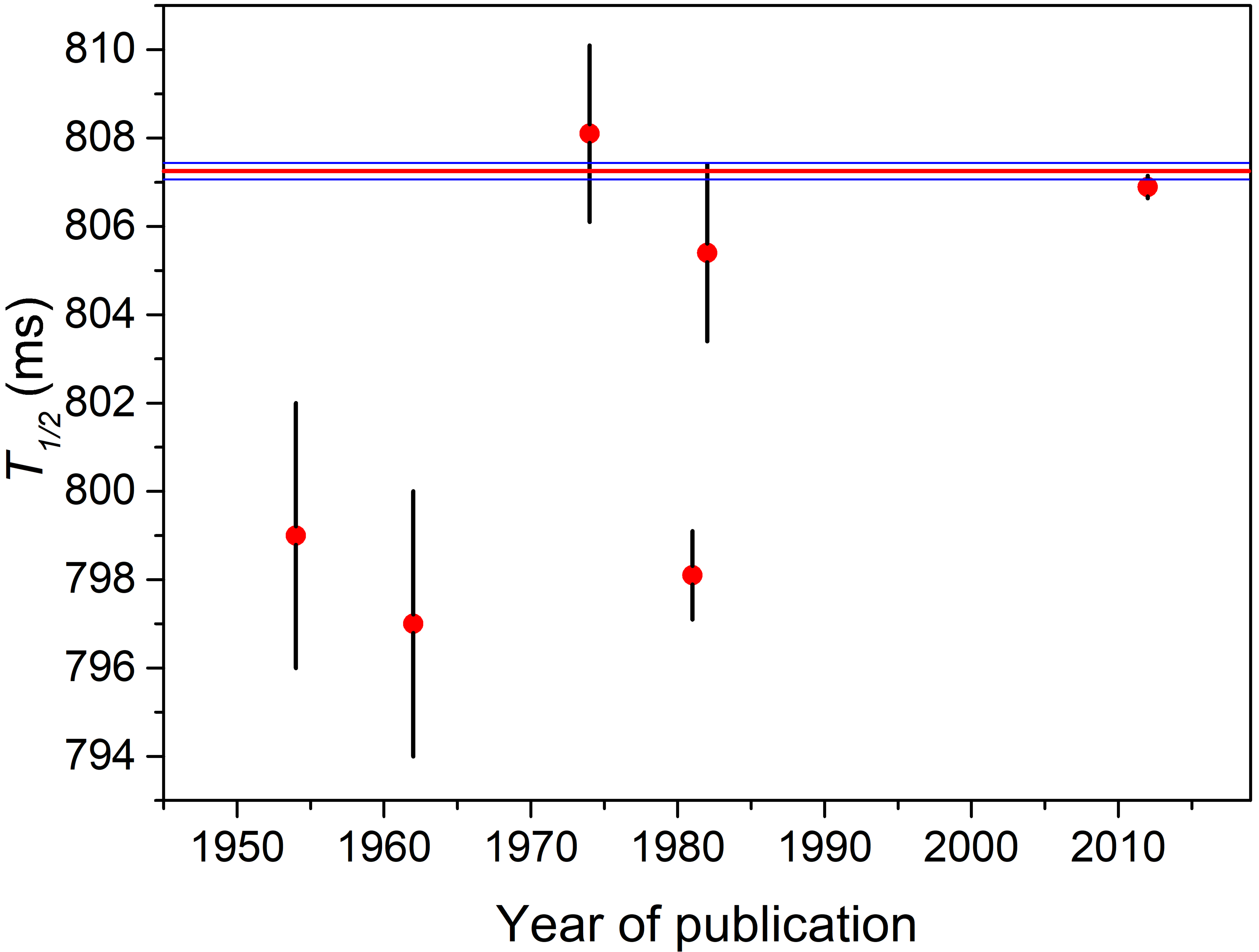}
\caption{Comparison between the $^6$He half-life value obtained from the present work (horizontal lines) and previous measurements having a relative precision smaller than 1\%. The plotted values are from Refs.\cite{Kli54,Bie62,Wil74,Bar81,Alb82,Kne12b}.}
\label{fig:t12-vs-year}
\end{figure}

Table \ref{tab:T12Values} gives the summary of the values obtained from the fits which include the main systematic corrections, whereas Table \ref{tab:systematics} gives the size of the systematic corrections to those values along with their associated uncertainties. The values from the three sets are statistically consistent, with differences in the central values smaller than one standard deviation. The systematic corrections add up to about 1~ms. The largest shift is due to the gain variation correction
and indicates how crucial it is to control this effect when aiming at a relative precision smaller than $10^{-3}$.
The combined results of the three measurement sets yield the value
\begin{equation}
\label{eq:T12}
T_{1/2} = (807.25\pm 0.16_{\rm stat}\pm 0.11_{\rm syst})~~{\rm ms}~,
\end{equation}
where the largest systematic uncertainty has been adopted. 
Figure~\ref{fig:t12-vs-year} shows a comparison between the present result (horizontal lines) and the six previously measured values having a relative precision smaller than 1\%. The present result is consistent with three of the previous values \cite{Wil74,Alb82,Kne12b} and
is at variance with three others \cite{Kli54,Bie62,Bar81}. Together with the result from Ref.~\cite{Kne12b}, which has a similar precision but used a different technique, the
present result confirms the $^6$He half-life close to 807~ms and strongly disfavors previous results yielding values below 800~ms.

\begin{table}[!hbt]
\caption{Corrections to the half-life (in ms) associated
with the main sources of systematic effects for the three sets of data.
The values of the uncertainties are rounded at 0.01~ms. The combined uncertainties of
the total correction were obtained by summation in quadrature.} 
\label{tab:systematics}
\begin{ruledtabular}
\begin{tabular}{lccc}
  Source                 & Set (1) 	 & Set (2)     & Set (3)\\
\colrule
Gain         &  0.75(7)       &     0.77(10)      &   0.78(6)\\
Baseline     &  0.09(3)       &     0.04(2)      &   0.05(9)\\
Pile-up      &  0.10(1)       &     0.25(1)      &   0.11(1)\\
Binning      &  $<0.01$    &     $<0.01$   &   $<0.01$\\
\colrule
Total correction  &  0.94(7)       &     1.06(11)      &   0.94(11)
\end{tabular}
\end{ruledtabular}
\end{table}

\section{CONCLUSION}
This work reported the most precise value of the $^{6}$He half-life obtained so far.
The result is consistent with the most recent measurement which supported two previous
values around 807~ms. By recording both the deposited energy and the time of each event with a digital data acquisition system combined with the use of a monitoring $^{241}$Am source, detector gain variations and dead-time effects were precisely measured and corrected for. These effects were found to contribute at 
a relative level of 10$^{-3}$ with a resulting relative systematic uncertainty at the level of 10$^{-4}$.

\section{ACKNOWLEDGMENTS}
The authors thank the LPC Caen and GANIL staffs for the technical support and
are grateful to A. Singh for her assistance during the running of the experiment.
This project was supported in part by the French Agence Nationale de la Recherche under grant ANR-20-CE31-0007-01 (bSTILED). 

%

\end{document}